\newcommand{\bnabla}{\mbox{\boldmath $\nabla$}}
\newcommand{\bcdot}{\mbox{\boldmath $\,\cdot\,$}}
\documentclass[aps,prb,twocolumn]{revtex4-1}
\usepackage{graphicx,amssymb,amsmath}
\begin{document}

\title{Magnetic dipole moment of a moving electric dipole}
\author{V. Hnizdo}
\affiliation{National Institute for Occupational Safety and Health,
Morgantown, West Virginia 26505}

\maketitle

The relativistic transformations of the polarization (electric moment density) $\bf P$ and magnetization (magnetic moment density) $\bf M$ of macroscopic electrodynamics\cite{PP} imply  corresponding transformations
of the electric and magnetic dipole moments $\bf p$ and $\bf m$, respectively, of a  particle. Thus, to first order in $v/c$,\cite{footnote1}
\begin{align}
{\bf p} &= {\bf p}_0 +\frac{1}{c}\,{\bf v}\times{\bf m}_0,
\label{tr_p}\\
{\bf m} &= {\bf m}_0 -\frac{1}{c}\,{\bf v}\times{\bf p}_0.
\label{tr_m}
\end{align}
Here, the subscripts 0 denote quantities in the particle's rest frame and $\bf v$ is the particle's velocity.
According to Eq.\ (\ref{tr_p}), a moving rest-frame magnetic dipole ${\bf m}_0$ develops an electric dipole moment ${\bf p}={\bf v}\times{\bf m}_0/c$. While this fact is
well known and understood,\cite{PP2,BK,Griffiths}  the complementary effect that a moving electric dipole acquires a magnetic moment does not seem to be understood equally well.\cite{footnote2}  There does not appear to be a consensus in the literature as to the transformation (\ref{tr_m}). For example,  Barut\cite{Barut} and Vekstein\cite{Vekstein} agree to first order in $v/c$ with (\ref{tr_m}), but  according to Fisher\cite{Fisher}
\begin{equation}
{\bf m} = {\bf m}_0 -\frac{1}{2c}\,{\bf v}\times{\bf p}_0,
\label{Fisher_m}
\end{equation}
implying that a moving rest-frame electric dipole ${\bf p}_0$ acquires a magnetic dipole moment ${\bf m}=-{\bf v}\times{\bf p}_0/2c$, which
differs by a factor of 1/2 from that of Eq.\ (\ref{tr_m}). The tasks of some problems involving a moving electric dipole in the authoritative text of Jackson\cite{Jack} seem  at first sight to be consistent with Fisher's transformation (\ref{Fisher_m}).

The aim of this note is to clear up the inconsistency of the differing transformations (\ref{tr_m}) and (\ref{Fisher_m}). To this end, we shall first express the electric current created by a moving electric dipole as the sum of polarization and magnetization currents, and then calculate the magnetic field of the moving dipole as the sum of the magnetic fields due to these currents.

A rest-frame electric dipole ${\bf p}_0$, located at ${\bf r}={\bf r}_0(t)$ and assumed for simplicity to be a point-like particle,  produces a polarization
\begin{equation}
{\bf P}({\bf r},t)={\bf p}_0\,\delta({\bf r}-{\bf r}_0(t)).
\label{P}
\end{equation}
Although its net charge vanishes, the dipole is a carrier of a bound charge distribution
\begin{align}
\rho_b({\bf r},t)&=-\bnabla\bcdot{\bf P}({\bf r},t)\nonumber\\
&=-{\bf p}_0\bcdot\bnabla\delta({\bf r}-{\bf r}_0(t)).
\label{rho}
\end{align}
When the dipole is moving with a velocity ${\bf v}=d{\bf r}_0(t)/dt$, it creates a current density
\begin{align}
{\bf J}({\bf r},t)&={\bf v} \rho_b({\bf r},t)\nonumber\\
&=-{\bf v}({\bf p}_0\bcdot\bnabla)\delta({\bf r}-{\bf r}_0(t)).
\label{J}
\end{align}
The same current density can obviously be obtained by modeling the moving electric dipole
as two equal and opposite point charges, separated by an infinitesimal displacement and moving with the same velocity $\bf v$.

On the other hand, the time-dependent polarization (\ref{P}) produces a polarization current density ${\bf J}_p({\bf r},t)$ according to
\begin{align}
{\bf J}_p({\bf r},t)&= \frac{\partial{\bf P}({\bf r},t)}{\partial t}\nonumber \\
&=-{\bf p}_0({\bf v}\bcdot\bnabla) \delta({\bf r}-{\bf r}_0(t)),
\label{Jp}
\end{align}
assuming that the dipole moment ${\bf p}_0$ itself does not depend on time.\cite{pdot}
If the moving electric dipole develops a magnetic dipole moment $\bf m$, then, in addition to the polarization (\ref{P}), there is also a magnetization
\begin{equation}
{\bf M}({\bf r},t)={\bf m}\,\delta({\bf r}-{\bf r}_0(t)),
\label{M}
\end{equation}
which produces a magnetization current ${\bf J}_m({\bf r},t)$ according to
\begin{align}
{\bf J}_m({\bf r},t)&= c\bnabla\times {\bf M} \nonumber \\
&=c\bnabla\times[{\bf m}\,\delta({\bf r}-{\bf r}_0(t))].
\label{Jm}
\end{align}
If now the magnetic dipole moment $\bf m$ generated by the motion of the electric dipole is given by
\begin{equation}
{\bf m}=-\frac{1}{c}\,{\bf v}\times{\bf p}_0,
\label{m}
\end{equation}
then the sum of the bound currents (\ref{Jp}) and (\ref{Jm}) equals
the ``convection" current (\ref{J}). Indeed, we then have
\begin{align}
&{\bf J}_p({\bf r},t)+{\bf J}_m({\bf r},t)\nonumber\\
&\;\:=-{\bf p}_0({\bf v}\bcdot\bnabla) \delta({\bf r}-{\bf r}_0(t))
-\bnabla\times[({\bf v}\times{\bf p}_0)\delta({\bf r}-{\bf r}_0(t))]\nonumber \\
&\;\:=-{\bf v}({\bf p}_0\bcdot\bnabla)\delta({\bf r}-{\bf r}_0(t)),
\label{JpJm}
\end{align}
where the last line was obtained using standard vector calculus identities.
The circulating magnetization current due to the magnetic moment ({\ref m}) is modified by the currents arising from the polarization current density (\ref{Jp}) so that the resulting net current is directed along the dipole's velocity $\bf v$.

In the nonrelativistic (quasi-static) limit, the vector potential due to the polarization current (\ref{Jp}) is
\begin{align}
{\bf A}_p({\bf r},t) &=\frac{1}{c}\int d^3r'\,\frac{{\bf J}_p({\bf r}',t)}{|{\bf r}-{\bf r}'|}\nonumber\\
&=-\frac{{\bf p}_0}{c}\int d^3r'\frac{({\bf v}\bcdot\bnabla') \delta({\bf r}'-{\bf r}_0(t))}{|{\bf r}-{\bf r}'|}\nonumber\\
&=\frac{{\bf p_0}}{c}\int d^3r'\delta({\bf r'}-{\bf r}_0(t))\frac{{\bf v}\bcdot({\bf r}-{\bf r}')}{|{\bf r}-{\bf r}'|^3}\nonumber\\
&=\frac{1}{c}\,\frac{{\bf v}\bcdot({\bf r}-{\bf r}_0(t))}{|{\bf r}-{\bf r}_0(t)|^3}\,{\bf p}_0.
\label{Ap}
\end{align}
Here, the integral in the 2nd line was performed by parts and the integral of the resulting gradient term was transformed into a vanishing surface integral.
The vector potential of the magnetization current (\ref{Jm}) that is due to the magnetic moment (\ref{m}) is evaluated similarly:
\begin{align}
{\bf A}_m({\bf r},t)&=\frac{1}{c}\int d^3r'\,\frac{{\bf J}_m({\bf r}',t)}{|{\bf r}-{\bf r}'|}\nonumber\\
&=-\frac{1}{c}\int d^3r'\,\frac{\bnabla'\times[({\bf v}\times {\bf p}_0)\,\delta({\bf r}'-{\bf r}_0(t))]}{|{\bf r}-{\bf r}'|}\nonumber\\
&=-\frac{1}{c}\,\frac{({\bf v}\times{\bf p}_0)\times({\bf r}-{\bf r}_0(t))}{|{\bf r}-{\bf r}_0(t)|^3}.
\label{Am}
\end{align}
The magnetic field  ${\bf B}$ of the moving electric dipole is therefore the sum
\begin{equation}
{\bf B}={\bf B}_m+{\bf B}_p,
\label{Bmp}
\end{equation}
where
\begin{align}
{\bf B}_m&=\bnabla\times {\bf A}_m({\bf r},t)\nonumber\\
&=-\frac{1}{c}\,\frac{3[{\bf n}\bcdot({\bf v}\times{\bf p}_0)]{\bf n}{-}{\bf v}\times{\bf p}_0}{|{\bf r}-{\bf r}_0(t)|^3},
\label{BM}
\end{align}
the magnetic field due to the vector potential (\ref{Am}),
is the magnetic field of the magnetic moment (\ref{m}), and
\begin{align}
{\bf B}_p&=\bnabla\times {\bf A}_p({\bf r},t)\nonumber\\
&=\frac{1}{c}\,{\bf p}_0\times\frac{3({\bf n}\bcdot{\bf v}){\bf n}-{\bf v}}{|{\bf r}-{\bf r}_0(t)|^3}
\label{BP}
\end{align}
is the magnetic field due to the vector potential (\ref{Ap}), created by the polarization current (\ref{Jp}). Here,
\begin{equation}
{\bf n}=\frac{{\bf r}-{\bf r}_0(t)}{|{\bf r}-{\bf r}_0(t)|}
\label{n}
\end{equation}
is a unit vector along the direction from the dipole's location ${\bf r}_0$ to the field point $\bf r$.
The sum of the expressions (\ref{BP}) and (\ref{BM}) reduces to
\begin{equation}
{\bf B} = \frac{1}{c}\,{\bf v}\times {\bf E},
\label{B}
\end{equation}
where
\begin{equation}
{\bf E} = \frac{3({\bf n}\bcdot{\bf p}_0){\bf n}-{\bf p}_0}{|{\bf r}-{\bf r}_0(t)|^3}
\label{Ep}
\end{equation}
is the electric field of the moving dipole in the nonrelativistic limit.
We note that the magnetic field (\ref{B}) is the same as that
obtained by Lorentz transforming to first order in $v/c$  the dipole's
rest-frame electromagnetic field to the laboratory frame.

The dipole transformation (\ref{tr_m}), implying that a
moving rest-frame electric dipole ${\bf p}_0$ acquires a magnetic dipole moment ${\bf m}=-{\bf v}\times{\bf p}_0/c$, is thus correct, though the magnetic field due to the dipole moment $\bf m$ is not the whole magnetic field of the moving electric dipole, which includes also the magnetic field created by the polarization current.\cite{VH}  The magnetic dipole moment $\frac{1}{2}{\bf m}=-(1/2c){\bf v}\times{\bf p}_0$ arises in a different decomposition of the moving dipole's magnetic field,\cite{Jack2}
\begin{equation}
{\bf B} = {\bf B}_{m/2}-\frac{3}{2c}\,{\bf n}\times \frac{({\bf v}\bcdot{\bf n})
{\bf p}_0+({\bf p}_0\bcdot{\bf n}){\bf v}}{|{\bf r}-{\bf r}_0(t)|^3},
\label{B_Jack}
\end{equation}
where the first term is the magnetic field due to the dipole moment $\frac{1}{2}{\bf m}$
and the second term,  which is symmetric in ${\bf p}_0$ and $\bf v$, has a curl that is
proportional to the displacement current of the electric quadrupole field that is created  when the electric dipole's location is off the origin, i.e. ${\bf r}_0\ne 0$.\cite{Jack2}

The transformation (\ref{Fisher_m}) of Fisher was arrived at using the standard definition ${\bf m} =(1/2c)\int d^3r\, {\bf r}\times {\bf J}$.
The origin of this definition is in a multipole expansion of the vector potential of a localized current distribution  that is assumed to be
divergenceless, i.e.\ $\bnabla\bcdot {\bf J}=0$, so that the value of $\bf m$ is independent of the choice of the reference point ${\bf r}=0$.\cite{Jack3}
However, the net current density (\ref{J}) of a moving electric dipole is not divergenceless: by the continuity equation, $\bnabla\bcdot {\bf J}=-\partial\rho_b/\partial t$, and $\partial\rho_b/\partial t\ne 0$ for the moving dipole's charge density $\rho_b$ (see Eq.\ (\ref{rho})).
But as a curl, the magnetization component ${\bf J}_m$ of the net current density {\bf J} is divergenceless. It is only the magnetization current ${\bf J}_m =-\bnabla\times[({\bf v}\times{\bf p_0})\delta({\bf r}-{\bf r}_0)]$ that, using  ${\bf m} =(1/2c)\int d^3r\, {\bf r}\times {\bf J}_m$, determines a magnetic dipole moment that is appropriate for a moving electric dipole.\cite{footnote3}

The full magnetic field of a moving electric dipole can be decomposed in more than one way. Decomposition (\ref{B_Jack}) is not incorrect, but decomposition (\ref{Bmp}) is singled out by the clearcut sources of its two terms: the divergenceless magnetization current due to the magnetization produced by the magnetic dipole moment ${\bf m}= -{\bf v}\times {\bf p}_0/c$ that is consistent with the transformation relations of special relativity and the non-zero-divergence polarization current due to the polarization  produced by a moving electric dipole.\cite{footnote4}

\section*{Acknowledgments}
The author thanks Grigory Vekstein for illuminating correspondence and comments, in particular for the insight that the standard definition of magnetic moment should not be used with a current distribution that is not divergenceless. Kirk McDonald and David Griffiths are thanked for helpful comments on drafts, and  J. David Jackson for illuminating correspondence.
This note was written by the author in his private capacity. No official support or endorsement by the Centers for Disease Control and Prevention is intended or should be inferred.


\begin{thebibliography}{99}
\bibitem{footnote1} We use Gaussian units since the relativistic aspects of relations among the dipole moments are displayed most clearly in these units.
\bibitem{footnote2}  This situation is somewhat paradoxical since we shall see that, unlike the acquired electric dipole moment, the acquired magnetic dipole moment is not a relativistic effect.
\bibitem{footnote3} It can be shown easily by integration by parts that $(1/2c)\int d^3r\, {\bf r}\times {\bf J}_m$, where  ${\bf J}_m = c\bnabla \times[{\bf m}\,\delta({\bf r}{-}{\bf r}_0)]$,  equals the magnetic moment $\bf m$ in the curl expression for ${\bf J}_m$.
\end{thebibliography}
\end{document}